\documentclass[conference]{IEEEtran}
\IEEEoverridecommandlockouts

\usepackage{cite}
\usepackage{amsmath,amssymb,amsfonts}
\usepackage{algorithmic}
\usepackage{graphicx}
\usepackage{textcomp}
\usepackage{xcolor}
\usepackage{enumitem}

\usepackage{xcolor}

\usepackage{url}
\usepackage{comment}
\usepackage{subfig}
\usepackage{multirow}
\def\BibTeX{{\rm B\kern-.05em{\sc i\kern-.025em b}\kern-.08em
    T\kern-.1667em\lower.7ex\hbox{E}\kern-.125emX}}
    
\begin{document}

\title{HotSwap: Enabling Live Dependency Sharing in Serverless Computing
\thanks{This work was accepted at the IEEE International Conference on Cloud Computing 2025.}
}
\author{
\IEEEauthorblockN{Rui Li}
\IEEEauthorblockA{
\textit{Northeastern University} \\
Boston, MA, USA \\
li.rui8@northeastern.edu}
\and
\IEEEauthorblockN{Devesh Tiwari}
\IEEEauthorblockA{
\textit{Northeastern University} \\
Boston, MA, USA \\
d.tiwari@northeastern.edu}
\and
\IEEEauthorblockN{Gene Cooperman}
\IEEEauthorblockA{
\textit{Northeastern University} \\
Boston, MA, USA \\
gene@ccs.neu.edu}
}

\maketitle

\begin{abstract}
This work presents HotSwap, a novel provider-side cold-start optimization for serverless computing. 
This optimization reduces cold-start time when booting and loading dependencies at runtime inside a function container. 
Previous research has extensively focused on reducing cold-start latency for specific functions.
However, little attention has been given to skewed production workloads.
In such cases, cross-function optimization becomes essential.
Without cross-function optimization, a cloud provider is left with two equally poor options:
(i) Either the cloud provider gives up optimization for each function in the long tail (which is slow); or
(ii) the cloud provider applies function-specific optimizations (e.g., cache function images) to every function in the long tail (which violates the vendor's cache constraints).
HotSwap demonstrates cross-function optimization using a novel pre-warming strategy.
In this strategy, a pre-initialized live dependency image is migrated to the new function instance. 
At the same time, HotSwap respects the provider's cache constraints, because a single pre-warmed dependency image in the cache can be shared among all serverless functions that require that image.
HotSwap has been tested on seven representative functions from FunctionBench.
In those tests, HotSwap accelerates dependency loading for those serverless functions with large dependency requirements by a factor ranging from 2.2 to 3.2.
Simulation experiments using Azure traces indicate that HotSwap can save 88\% of space, compared with a previous function-specific method, PreBaking, when sharing a dependency image among ten different functions.
\end{abstract}


\section{Introduction}
\label{sec:intro}
Serverless computing is a flexible and popular cloud computing methodology that provides users with a simpler and cheaper way to deploy applications. 
A common example is Function-as-a-Service (FaaS), which enables event-driven execution of individual functions.
It has quickly gained attention across various cloud providers, including AWS~\cite{AWSLambda}, Azure~\cite{AzureFunction}, Google\cite{googlecloud}, and Alibaba~\cite{AlibabaFunction}. 

However, many of the financial benefits are undermined by the \emph{cold-start} problem. 
In FaaS, the cloud provider loads user function into the serverless instance only upon user invocation.
After an invocation, the cloud retains the serverless instance in memory for a certain period (i.e., during a \emph{keep-alive time}).
If an invocation occurs when no instance is alive in the system, the function experiences a \emph{cold start}, which is significantly slower due to instance initialization. 
Otherwise, the function experiences a faster \emph{warm start} with no initialization phase.
Due to the significance and challenges of the cold start problem, it has been extensively studied by systems researchers~\cite{ristov2022colder, manner2018cold, ebrahimi2024cold}.

A naive solution to avoid the \emph{cold start} is to always keep the function instance alive, ensuring that cold starts never occur. 
Researchers have further developed approaches to intelligently adjust the keep-alive duration based on invocation patterns~\cite{fuerst2021faascache,shen2021defuse} or by predicting invocation arrivals and pre-loading functions in advance~\cite{roy2022icebreaker, fuerst2021faascache}. However, most user request arrivals are hard to predict~\cite{shahrad2020serverless, wang2018peeking}.

Keeping idle functions alive wastes resources, prompting research into accelerating cold starts instead.  
A common approach is to pre-warm functions using checkpoint images or caches~\cite{silva2020prebaking, fireman2024prebaking, du2020catalyzer, cadden2020seuss}, which bypass parts of the initialization process.  
While effective, these methods increase memory usage, straining limited cache resources.  
Recent work refines this by pre-fetching hot pages~\cite{ustiugov2021benchmarking} and generating function seeds~\cite{wei2023no}, but further optimization is needed to reduce overhead and improve performance.

Experiments from a recent state-of-the-art work show that these methods improve the cold-start tail latency by 89\% and require only 29MB of memory per machine to pre-warm a serverless function~\cite{wei2023no}. 
These methods optimize cold starts by analyzing function-specific initialization bottlenecks.  
However, they are not lightweight, often requiring deep analysis or specialized OS and hardware support~\cite{wei2023no}, such as RDMA.

Nevertheless, according to Azure Function traces~\cite{shahrad2020serverless}, 50\% of functions experience fewer than $1.4$ cold starts per day, assuming a keep-alive time of 15 minutes. 
Therefore, maintaining pre-initialized images or recording hot pages in RAM for every serverless function in a workload like Azure~\cite{shahrad2020serverless} is costly and unnecessary.

\begin{figure}[t] 
  \includegraphics[width=0.48\textwidth]{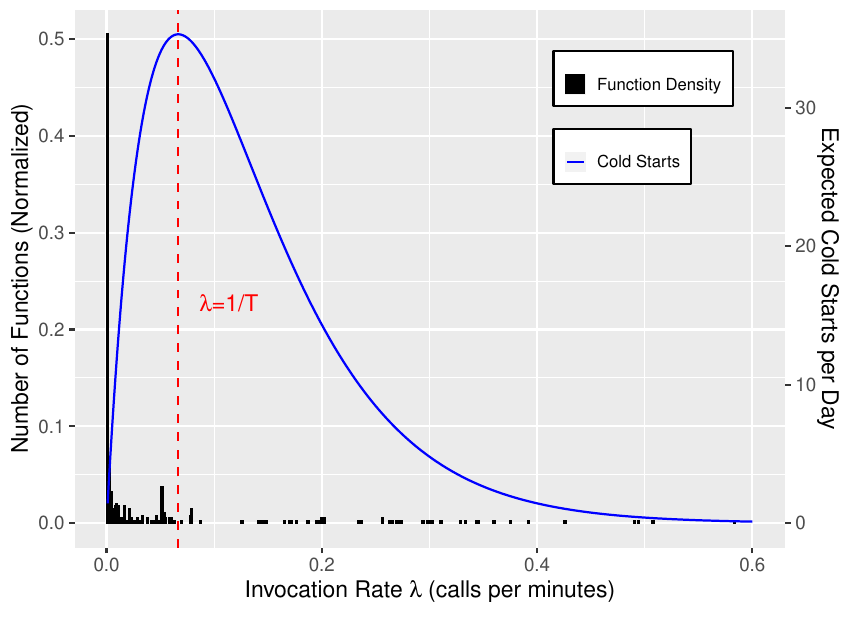}
  \caption{The relationship between invocation rate $\lambda$ and expected cold starts with keep-alive time $T=15$~minutes, using function distribution from Azure~\cite{shahrad2020serverless}. Each serverless function was assigned to a bucket $[x, x+0.001]$, where the $\lambda$ of the serverless function was between $x$ and $x+0.001$ invocations per minute. The y-value for a bucket is the normalized number of functions in that bucket.}
  \label{fig:invocation_rate_and_cold_start}
\end{figure}

Implementing function-specific optimizations for all serverless functions is impractical in production due to skewed workloads, where many functions experience few cold starts (see Figure~\ref{fig:invocation_rate_and_cold_start}).  
For functions with invocation rates far from $1/T$, the cost of applying prior methods is not economical, where $T$ is the keep-alive time. 

A more practical approach, albeit with other limitations, is \emph{cross-function pre-warming}, where shared resources are preserved across functions~\cite{oakes2018sock, huang2024trenv, lin2021flashcube}.  
For instance, SOCK~\cite{oakes2018sock} and FlashCube~\cite{lin2021flashcube} reduce startup latency respectively by retaining Zygote containers or by retaining shared components on worker nodes.  
However, maintaining or migrating such pre-initialized containers remains costly at scale.

\textbf{Despite extensive research on the cold start problem, key challenges remain and hinder large-scale deployment in production.}
In particular, there is a lack of lightweight, cost-efficient, cross-function optimizations that target the majority of infrequently invoked functions (see the caption of Figure~\ref{fig:invocation_rate_and_cold_start}).
This paper aims to enable \emph{cross-function} optimizations in the dependency-loading phase, a major bottleneck in current production cloud environments (see Section~\ref{sec:AWS_performance_analysis}), while keeping overhead minimal.  Thus, we need only apply cross-function optimization to the frequently used dependencies instead of entire images, thereby improving scalability.

\subsection*{The HotSwap Approach:}
In this paper, we present \textit{HotSwap}, a cross-function solution targeting the dependency-loading overhead during cold starts. 
HotSwap enables the provider to maintain a pool of \emph{live dependency images} in RAM. 
A live dependency image is a memory-resident checkpoint representing software dependencies. 
Typically, these images include popular software dependency combinations, such as NumPy and Torch.

HotSwap is designed to benefit serverless providers by also optimizing the massive number of less frequently invoked functions (the ``long tail'' of the distribution), which is usually not practical in a function-specific approach.
HotSwap brings an important benefit. 
Providers only need to maintain a limited number of live dependency images in memory, which can be shared across a wide range of serverless functions using that dependency.
Therefore, for functions sharing a common dependency, HotSwap represents a new type of cross-function optimization using pre-warmed dependency images, offering several advantages over previous approaches in the literature.
Supporting this, a study~\cite{oakes2018sock} analyzed 876,000 Python projects and found that just 20 packages accounted for 36\% of all `import' statements, indicating that most functions rely on a small set of common dependencies.

A prototype of HotSwap was evaluated by running offline AWS Lambda containers on AWS EC2 nodes.
The implementation of HotSwap, unlike some other approaches, is entirely transparent, without modification to the operating system kernel or the source code of the serverless function. 
In our tests, loading libraries such as ``sklearn + pandas'', ``NumPy + keras'', and ``NumPy + torch'' through HotSwap was up to 2.2x, 3.2x, and 2.5x faster, respectively, compared to traditional AWS Lambda startup processes. 
Furthermore, experiments in Section~\ref{sec:sharing_exp} using traces from Azure~\cite{shahrad2020serverless} show that HotSwap can save 88.4\% of memory space compared to a function-specific approach, PreBaking~\cite{silva2020prebaking, fireman2024prebaking}. 
This efficiency is achieved when warming up 10 different serverless functions that share the same dependency.

We summarize our contributions as follows.
\begin{enumerate}
\item
This paper proposes HotSwap, a novel cold-start optimization targeting skewed, long-tail serverless workloads, which are common in commercial clouds like Azure.
\item 
HotSwap introduces a new solution to the dependency-loading issue, where the provider stores middleware once in a separate dependency image, independent of the entire container. A single pre-warmed dependency image can be shared across different function instances.
\item 
HotSwap was evaluated for its ability to reduce provider-side costs (CPU usage, cache, RAM) during cold starts using a real-world workload from Azure~\cite{shahrad2020serverless}, rather than a synthetic workload.
\end{enumerate}

The rest of this paper is organized as: 
Section~\ref{sec:background} presents the background and motivation of our paper;
Section~\ref{sec:Method} presents the design and implementation details of HotSwap;
In Section ~\ref{sec:evaluation}, we evaluate the performance of HotSwap.
We then discuss related work in Section~\ref{sec:related_work}, and conclude in Section~\ref{sec:conclusion}.

\begin{figure*}[htb]
  \includegraphics[width=\textwidth]{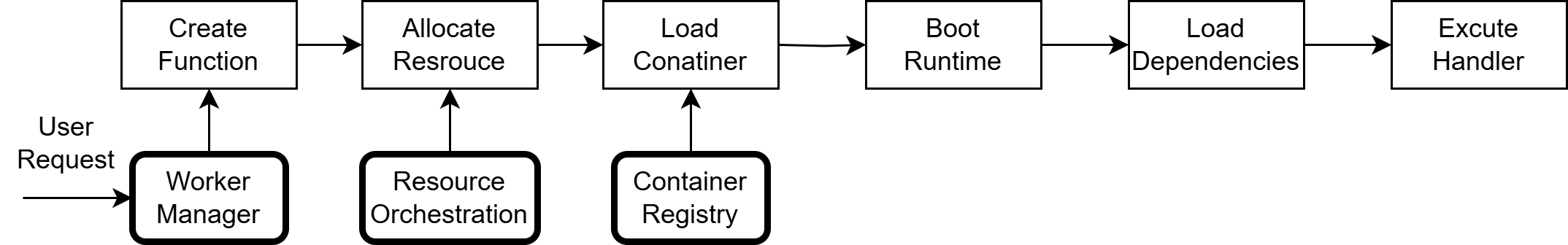}
  \caption{The process of serverless cold starts in the current production cloud}
  \label{fig:cold_start}
\end{figure*}

\section{Background and Motivation}
\label{sec:background}
This section introduces: the background for serverless computing and cold starts (Section~\ref{sec:serverless_background}); 
the need for HotSwap (Section~\ref{sec:invocation_pattern_vs_cold_start}); and
a performance bottleneck in a commercial cloud platform such as AWS Lambda
(Section~\ref{sec:AWS_performance_analysis}).

\subsection{Cold Start in Serverless Computing} 
\label{sec:serverless_background} 
This section introduces the background of container-based serverless computing from a cold-start perspective.
Figure~\ref{fig:cold_start} shows the general process of such a cold start.

In Figure~\ref{fig:cold_start}, while the specifics may differ across platforms, most serverless architectures consist of two core components: a \emph{Worker Manager} and a \emph{Resource Orchestration} unit.
The \emph{Worker Manager} responds to a request for capacity from the serverless front end.
The \emph{Worker Manager} acts as a stateful load balancer by continuously monitoring the capacity of each unique function on the platform.
If sufficient capacity exists (e.g., idle function instances are available), the \emph{Worker Manager} directs the front end to route the request to an idle instance.
Conversely, if resources are unavailable (triggering a cold start), it instructs the \emph{Resource Orchestration} unit to provision a new function instance with adequate resources.

Recall also that the user has uploaded the container function to the provider's \emph{Container Registry}.
Next, the provider will download the user container from the \emph{Container Registry} and start it.
The container initiates the \emph{serverless runtime} to boot the entire serverless function.
Then, the serverless runtime starts a specific function runtime (e.g., Python), and the program loads the necessary software dependencies (e.g., NumPy) and the user-specific \emph{handler code}.
The handler code is then ready to handle a new function instance.

To minimize the cold-start overhead, one needs to skip or expedite one or more steps.
When designing a solution for a specific serverless function, all function-specific steps can be considered. 
Previous work has designed end-to-end, function-specific optimization for targeted serverless functions and achieved substantial reductions in cold-start latency.
However, when performing cross-function optimization for the entire system, one can only optimize steps that are identical or similar across all serverless functions, such as container storage and loading.

In this paper, we explore how runtime booting and dependency loading can benefit from cross-function optimization, as they are common processes shared by multiple serverless functions.
Thus, we propose HotSwap as a novel cross-function optimization aimed at reducing the time required for runtime booting and dependency loading.

\subsection{Impact of Invocation Patterns on Cold Starts}
\label{sec:invocation_pattern_vs_cold_start}
In this section, we aim to address two questions: 1) How frequently does a cold start occur for a serverless function? and 2) How does a skewed workload impact the optimization choices of the cloud provider?

To simplify the discussion, we focus on situations where a cold start occurs due to the absence of any active instances, as this scenario accounts for more than $99\%$ of cold starts in production systems~\cite{shahrad2020serverless}.

To address question~1, we model the occurrence of cold starts using an exponential distribution. 
The frequency of cold starts for a serverless function depends on its invocation pattern. 
For extreme cases, a function that is never invoked or invoked very frequently (e.g., every second) would not experience cold starts.
Let $\lambda$ denote the invocation rate (calls per minute), and $T$ denote the keep-alive time. The probability of no invocation occurring during the keep-alive time follows:
\begin{equation}
\label{eq
}
P_{no\_inv} = e^{-\lambda T}
\end{equation}
$P_{no\_inv}$ represents the probability that no invocation happens during the keep-alive timeout, leading to a potential cold start.

The number of expected cold starts that occur within $D$ minutes can be expressed as:
\begin{equation}
\label{eq:expect}
E_{cs}(\lambda)=D \lambda P_{no\_inv} = D\lambda e^{-\lambda T}
\end{equation}
which has a maximum when $d E/ d\lambda = 0, \lambda=1/T$. 

Figure~\ref{fig:invocation_rate_and_cold_start} shows the relation of $E_{cs}$ versus $\lambda$ when $T=15\;min$ and $D = 1440\;min$ (a day).
Only functions with invocation rates around $1/T$ have a large number of expected cold starts, functions that are less or more frequently invoked have fewer cold starts.
Cloud providers like Azure~\cite{AzureFunction} and AWS~\cite{AWSLambda} configure dynamic \emph{keep-alive} time from $5$ to $30$ minutes. 
The different \emph{keep-alive} time changes the position of the maximum point, but the shapes are similar.

To answer question 2, we draw the density histogram of the actual function distribution in the same Figure~\ref{fig:invocation_rate_and_cold_start}.
For function-specific tuning, assuming the cost of optimizing a function is~$c$ and the benefits of an optimized cold start are $w$.
Functions that are worth function-specific tuning must satisfy the following:
\begin{equation}
\label{eq:cost}
w E_{cs}(\lambda) > c
\end{equation}

However, from the histogram, the serverless workload in production can be extremely skewed.
More than $50\%$ of the functions have a very low invocation rate (less than 0.001 calls per minute) and $E_{cs}$, which experience less than $1.4$ times of cold starts per day.
Thus, those functions do not satisfy Equation~\ref{eq:cost}, which means function-specific tuning is not economical for them.

For functions whose invocation rates do not meet the tuning criterion in Equation~\ref{eq:cost}, only cross-function optimizations are feasible.  
One successful example is AWS Lambda~\cite{brooker2023demand}, which employs a three-tier caching system and container deduplication to load a 10\,GB serverless container in just 50\,ms.  
Such optimizations are applicable to all serverless functions, regardless of invocation frequency.  

In this paper, we aim for HotSwap to further enhance cross-function optimization, benefiting serverless functions where function-specific tuning is not economically viable.

\subsection{Performance Analysis of the Cold Start for Current AWS Lambda Functions}
\label{sec:AWS_performance_analysis}  

\begin{figure}[t]
    \centering
    \includegraphics[width=0.48\textwidth]{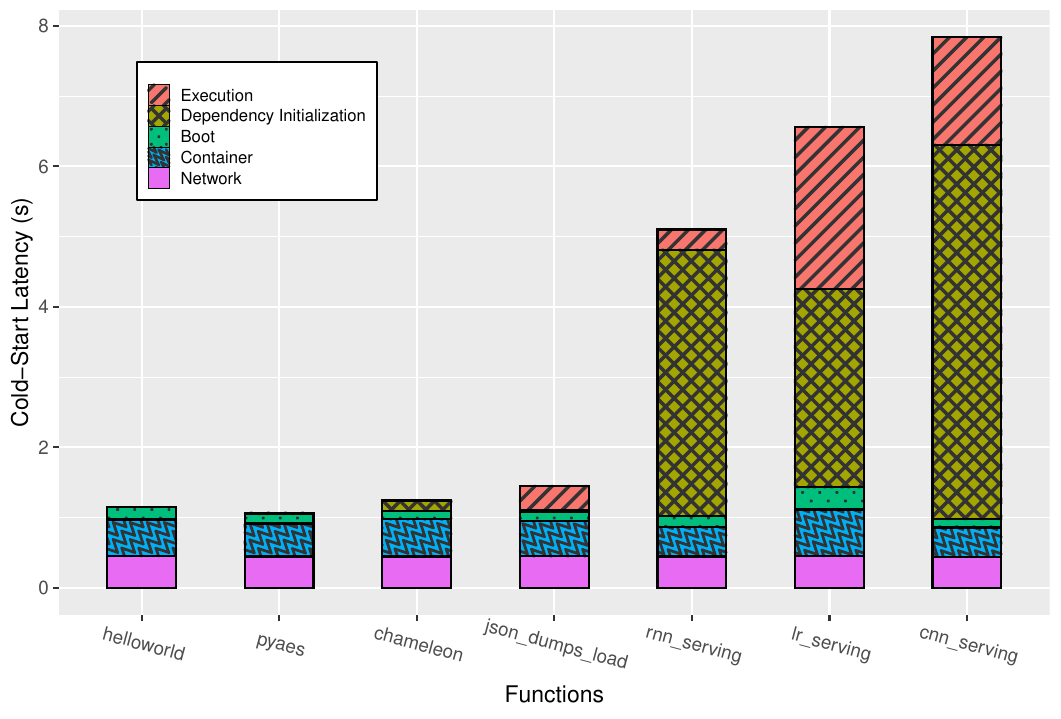}
    \caption{Cold-start latency breakdown for serverless functions in a traditional AWS Lambda function (not HotSwap).} 
    \label{fig:AWS_Performance}
\end{figure}

This section analyzes the performance of cold-start overhead in a production cloud on AWS~\cite{agache2020firecracker,brooker2023demand}.
The goal is to analyze which components of the current AWS Lambda architecture still have room for optimization.

For analysis, we build serverless function containers to create AWS Lambda functions, and then invoke them by a script.
All Lambda functions are configured with a 10~GB memory limitation to maximize the CPU resources.
To collect cold-start times, we delete and recreate lambda functions after each run.
There are several timers in the analysis script and the function containers to track the time of each cold-start process.
The entire time for the cold start is calculated from the script that sends the request to the serverless function and receives the result.

Note that the caching system of AWS Lambda would cause bias in a naive experimental design.
For example, if the user uploads a new function container, the first cold start will be extremely slow as compared to the following ones.
Moreover, if the functions are frequently invoked, AWS upgrades the function to a different cache that will improve its cold starts. (See ``AWS three-tier caching'' in~\cite{brooker2023demand}.)
The improved caching scheme of AWS Lambda continues to be used even when the function is deleted and then recreated.
In order to achieve reproducible numbers with lower variance in our own experiments on AWS Lambda, we upload a container and then create and delete a Lambda function many times, until the cold-start results are stable.

\begin{table}[ht] 
\centering
\caption{Serverless functions adopted from FunctionBench~\cite{kim2019functionbench}.}
\label{tab:AWS_performance}
\begin{tabular}{llll}
\hline
\multicolumn{1}{c}{\multirow{2}{*}{\begin{tabular}[c]{@{}c@{}}Serverless \\ Functions\end{tabular}}} & \multicolumn{1}{c}{\multirow{2}{*}{\begin{tabular}[c]{@{}c@{}}Container \\ Size (MB)\end{tabular}}} & \multicolumn{1}{c}{\multirow{2}{*}{\begin{tabular}[c]{@{}c@{}}Major\\ Dependencies\end{tabular}}} & \multicolumn{1}{c}{\multirow{2}{*}{\begin{tabular}[c]{@{}c@{}}Function\\ Purpose\end{tabular}}} \\
\multicolumn{1}{c}{}                                                                                 & \multicolumn{1}{c}{}                                                                                & \multicolumn{1}{c}{}        
 & \multicolumn{1}{c}{}     
 \\ \hline
helloworld                                                                                           & 180.00                                                                                              & None        
& Minimal function
\\
json\_dumps\_load                                                                                    & 180.00                                                                                              & urllib, json   
& Json Serialization
\\
pyase                                                                                                & 180.11                                                                                              & pyaes, string 
& Text encryption
\\ 
chameleon                                                                                            & 180.42                                                                                              & chameleon                                                                                         & HTML rendering
\\
lr\_serving                                                                                          & 379.52                                                                                              & sklearn, pandas                                                                                   & Logistic regressions
\\
cnn\_serving                                                                                         & 1386.73                                                                                             & NumPy, keras                                                                                      & Image classification 
\\
rnn\_serving                                                                                         & 5602.46                                                                                             & NumPy, torch                                                                                      & Sequence generation 
\\ \hline
\end{tabular}
\end{table}

Table~\ref{tab:AWS_performance} presents serverless functions from a representative serverless benchmark suite called FunctionBench~\cite{kim2019functionbench,kim2019practical}.
We have chosen FunctionBench for ease of comparison with previous work~\cite{ustiugov2021benchmarking}.
All these functions are written in Python and they use some representative Python packages like ``NumPy'', ``torch'', ``keras'', and ``sklearn''.
Table~\ref{tab:AWS_performance} lists both packages and the size of specific function container images.
Installing large packages also significantly increases the image size of the function containers. 

Figure~\ref{fig:AWS_Performance} shows the breakdown of intermediate times for cold starts in this analysis.
The entire cold-start time can be broken down into the following parts:
\begin{itemize}
    \item \emph{Network:} The process of network communications between invoker and user functions.
    \item \emph{Container:} The process of creating the AWS containers.
    \item \emph{Boot:} The process of booting Python and the AWS Lambda interface~\cite{AWS_lambda_runtime}.
    \item \emph{Dependency Initialization:} The process of importing required middleware, referred to as \emph{serverless software dependencies}.
    \item \emph{Execution:} The process of execution of user handler.
\end{itemize}

From Figure~\ref{fig:AWS_Performance}, the dependency initialization dominates the cold start of AWS Lambda functions.
In ``lr\_serving'', ``cnn\_serving'', and ``rnn\_serving'', the dependency initialization occupies $43\%$, $74\%$, and $68\%$ of the cold-start overhead, respectively.
Dependency initialization takes more time than execution for all three AWS Lambda functions.  
It remains a major bottleneck for serverless functions in production, especially since many rely on widely used packages.

Note that the time for container creation is almost constant (about~$0.5~s$), even for containers that have different image sizes. 
This shows that AWS has already optimized the time for the container creating, by improving the AWS caching and file systems~\cite{brooker2023demand,agache2020firecracker}.

In summary, the time for dependency initialization often dominates the cold-start time in AWS Lambda.
Therefore, improving the dependency initialization overhead is important both to serverless providers and to users of the current production cloud.
This motivates a novel method focusing on the dependency initialization overhead, as seen in Section~\ref{sec:Method}.

\begin{figure*}
  \centering 
  \subfloat[System Architecture]
  {
    \label{fig:Metholodgy_system_overview}\includegraphics[width=0.32\textwidth]{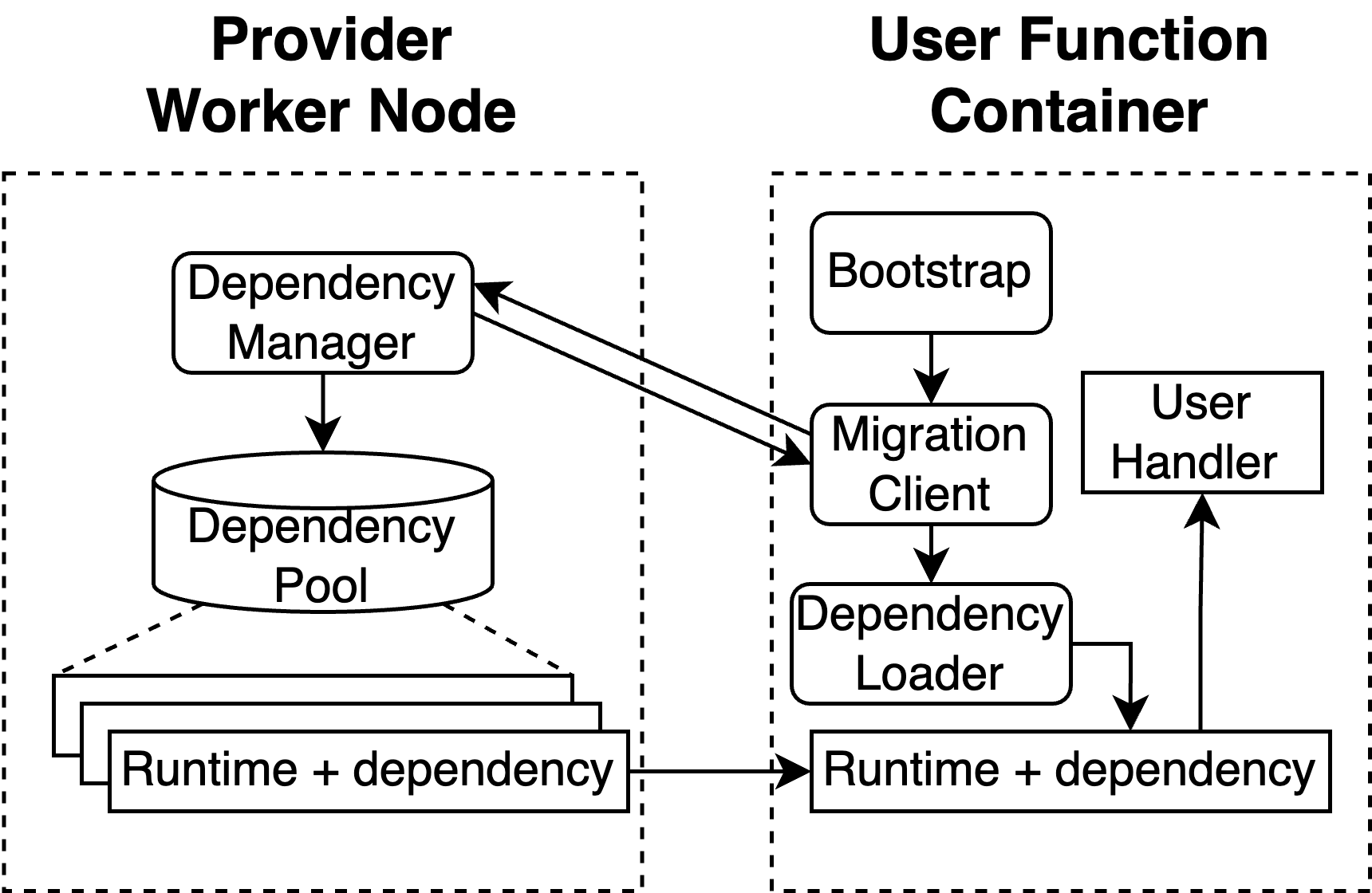}
  }
  \subfloat[Setup Phase]
  {
    \label{fig:Metholodgy_dependency_pool}\includegraphics[width=0.32\textwidth]{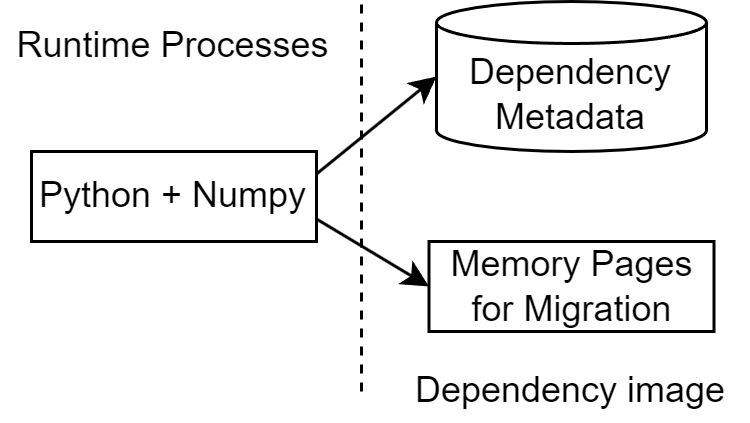}
  }
  \subfloat[Runtime Phase]
  {
    \label{fig:Metholodgy_lazy_migration}\includegraphics[width=0.32\textwidth]{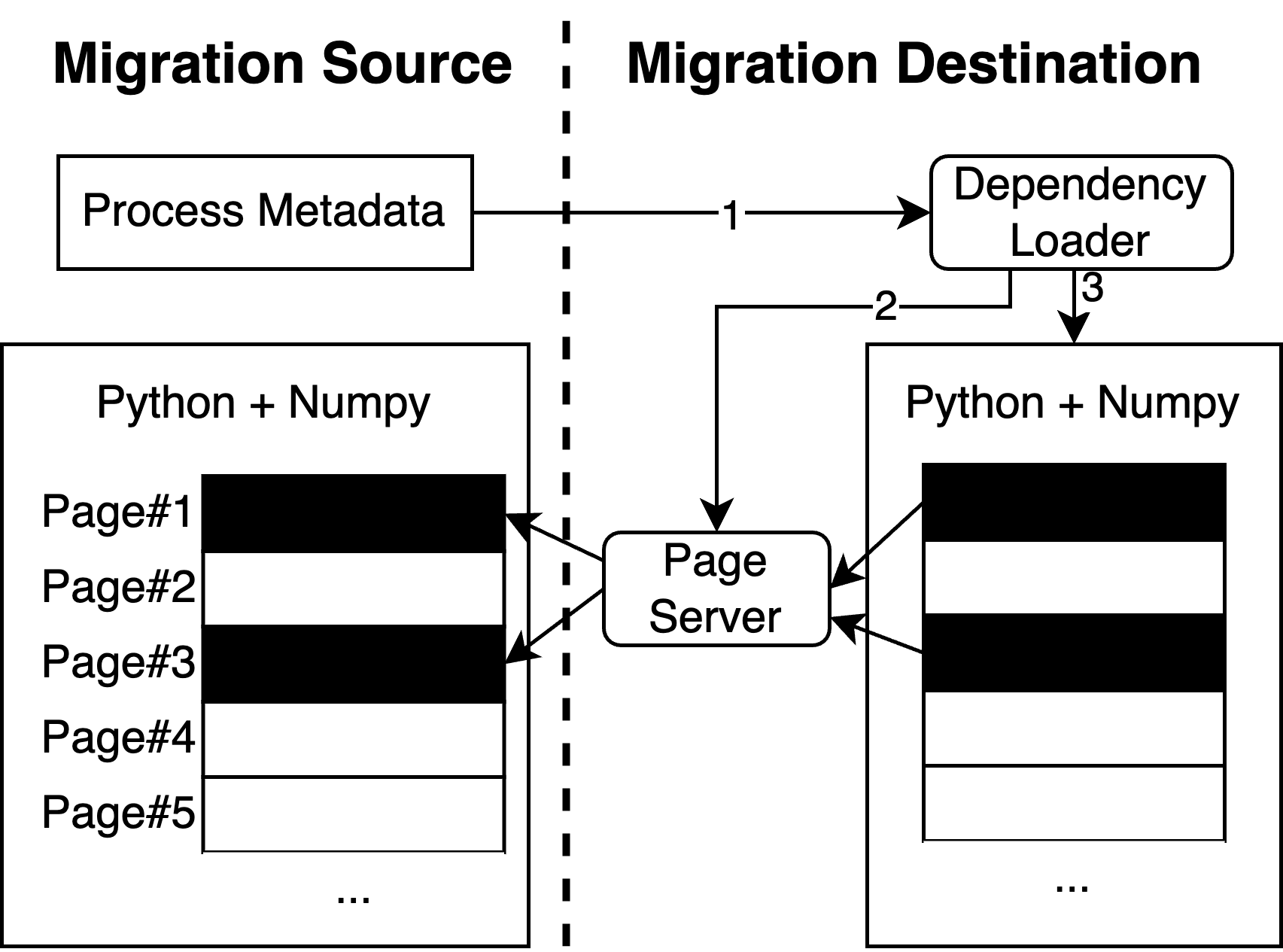} 
  }
  \caption{\textbf{System architecture and workflow:} (a)~Architectural overview of HotSwap and its dependency pool.
  (b)~In the setup phase, the user registers a serverless function, and HotSwap generates the corresponding image in the dependency pool. (c)~In the runtime phase, pages are copied from the image between provider worker nodes and user function containers.} 
  \label{fig:Metholodgy}
\end{figure*}

\section{HotSwap Design}
\label{sec:Method}  
This section introduces HotSwap, a prototype system for cross-function cold-start optimization.
HotSwap leverages the power of a shared dependency pool, enabling significant performance gains for providers.
A single global shared dependency pool for all users is implemented for cross-function optimization.
Hence, a single pre-initialized dependency image from the pool can benefit multiple users whose functions share a single common dependency.
In general, the pool contains several pre-initialized dependency images for popular runtimes and packages (e.g., NumPy, torch, etc.).
Thus, the cloud provider supports such a dependency pool for all users.

HotSwap enables a serverless function container to seamlessly migrate the dependency image of its \emph{pre-initialized} software dependencies from a provider's worker node.
For example, assume there is a serverless function based on NumPy. 
The function can use pre-initialized memory pages of NumPy through migration. 
This eliminates the otherwise large runtime overhead of first initializing NumPy.

\subsection{Motivation for the Design of HotSwap}
\label{sec:motivation}
HotSwap gains its performance advantages by exploiting three characteristics of the workloads seen by public cloud providers in their distribution of containerized serverless functions.
Those three characteristics are:

\begin{description}[style=unboxed,left=0pt]
\item[\textbf{Commonality:}]
Although there are numerous serverless software dependencies, and applications can use different combinations of dependencies, many applications are based on a small number of popular packages~\cite{oakes2018sock} (e.g., NumPy, torch). 

\item[\textbf{Sparsity:}]
Most applications do not use all functions from a package during execution~\cite{ustiugov2021benchmarking}.  
Thus, copying all memory pages to the function container beforehand is unnecessary.

\item[\textbf{Lightweight:}]
Keeping runtime and software dependencies in the memory is much more lightweight than keeping the whole container like~\cite{oakes2018sock}.
Thus, the provider can afford to maintain a pool of dependency images.
\end{description}

\subsection{Design Overview}
\label{sec:design_overview}
HotSwap is designed to reduce dependency loading overhead for serverless functions. 
The key idea is to \emph{migrate} pre-initialized dependencies instead of starting them from scratch. 
Here, \emph{migrate} means to copy a process from the host (which could be a local or remote node) to the user container over the network. 
The system design is based on the characteristics of such dependencies.
A container for a serverless function usually contains both user-specific code and its software dependencies.

Figure~\ref{fig:Metholodgy_system_overview} describes the components of HotSwap on the provider worker node and function container.
On the provider worker node, a \emph{Dependency Manager} orchestrates the management of in-memory \emph{dependency images}, which are images stored \emph{in memory} that represent a checkpoint of pre-initialized software dependencies. 
The Dependency Manager maintains metadata about existing live dependency images and manages the memory pages associated with them.
It is the central hub for communication with user function containers, facilitating migration services. 
When a migration request is initiated by a serverless container, the Dependency Manager accepts the request, transfers the essential metadata required for process restoration, and provides page transfer services to the serverless function container.

In the container for the user function, a \emph{Migration Client} orchestrates the migration of memory pages on the user side. 
Upon startup of the container, it initiates a migration request to the Dependency Manager. 
Upon receiving confirmation, it starts the \emph{Dependency Loader} to orchestrate the restoration of the runtime and the loading of dependency images, guided by the provided metadata. 

Rather than loading all memory pages immediately, the \emph{Dependency Loader} restores the memory pages of the dependency by \emph{lazy migration}.
In lazy migration, instead of the actual memory pages, the Dependency Loader initializes a \emph{Page Server} to handle this task later. When the user functions execute and access memory pages of the dependency, it triggers page faults, and the Page Server handles them. 
This Page Server communicates with the dependency pool to efficiently load memory pages.

\subsection{Generation of Dependency Images}
\label{sec:dependencies_pool}
Figure~\ref{fig:Metholodgy_dependency_pool} shows the setup phase of HotSwap, which dumps a serverless software dependency into a dependency image in the dependency pool. 
The serverless software dependency process in this example is a Python script importing NumPy and then executing the user handler.  
We initiate the dumping of the process after NumPy has been loaded but before the import and execution of the user code (the user handler in Figure~\ref{fig:Metholodgy_system_overview}).
In general, other shared dependency images might also be used besides NumPy.  The corresponding user handler will then execute immediately after that image has been migrated.

The dependency process is divided into two distinct components during the dumping process: (i)~process metadata, preserved within \emph{Dependency Metadata}; and (ii)~hot memory pages, residing securely within the \emph{Dependency Pool}.
The process metadata encompasses essential information for restoring the process framework, including file descriptors, memory segment sizes, and other relevant configuration details. 
The dependency pool preserves the full memory pages of the dependency process within its memory domain.
The process metadata, combined with its memory pages, is referred to as a \emph{dependency image} in this work.
To enable efficient page transfer, the Dependency Manager creates a source-side page server for each dependency image, which communicates over the network with destination page servers.
 
HotSwap requires only a modest pool of dependency images, because of the \emph{commonality} and \emph{lightweight} characteristics of serverless software dependencies.
Thus, cloud providers only undertake limited costs associated with dumping and maintaining dependency images.

\subsection{Migration of Dependency Images}
\label{sec:lazy_migration}
The migration client orchestrates the migration process, which has three distinct steps.

The first step of migration is to transport the process metadata to the destination (i.e., to the serverless function container).  
After receiving the migration request from the migration client, the dependency manager looks up the dependency metadata for the requested dependency image.
If the dependency image exists, the dependency manager transports process metadata to the dependency loader. 

The second step of migration is to start a \emph{page server}, which handles the ``page faults'' of the dependency to be loaded by using the ``userfaultfd'' feature of Linux.
The page server accesses the dependency pool through the network, to load those pages that are actually needed by the serverless function.

In the third step, the dependency loader then restores the serverless software dependencies based on the process metadata.
The dependency loader: (a)~restores the memory structure; and then (b)~reconnects file descriptors in the new environment.
After migration, the new process continues to import and execute the user handler.

For handling the ``page fault'', HotSwap could implement either of two possible policies:

\textbf{Lazy restore}: When handling the ``page fault'', only the pages required by the user functions are transferred. 
In this case, the Page Server needs to access the dependency pool every time a page fault occurs. 
Thus, the execution of the user function is slowed down by the number of pages the function needs. 
However, this method guarantees the transfer of only the minimum number of pages.

\textbf{Bulk restore}: When handling the first ``page fault'', the Page Server first transfers the required pages.
And then, in the background, it loads the remaining pages in parallel. 
In this case, the Page Server needs to access the dependency pool only once. 
Therefore, the execution will not be impacted by further page faults and remote communication. 
Although transferring all pages increases communication overhead, this should not impact the user functions because the Page Server can work in parallel with the user functions.

HotSwap chooses \emph{bulk restore} because it generally benefits all functions without impacting the warm starts (see Section~\ref{sec:variants_comp}).
In contrast, lazy restore works well only in more specialized page fault cases.

\begin{figure*}[ht]
  \centering 
  \subfloat[Cold Start]
  {
    \label{fig:com_bas_cold}\includegraphics[width=0.45\textwidth]{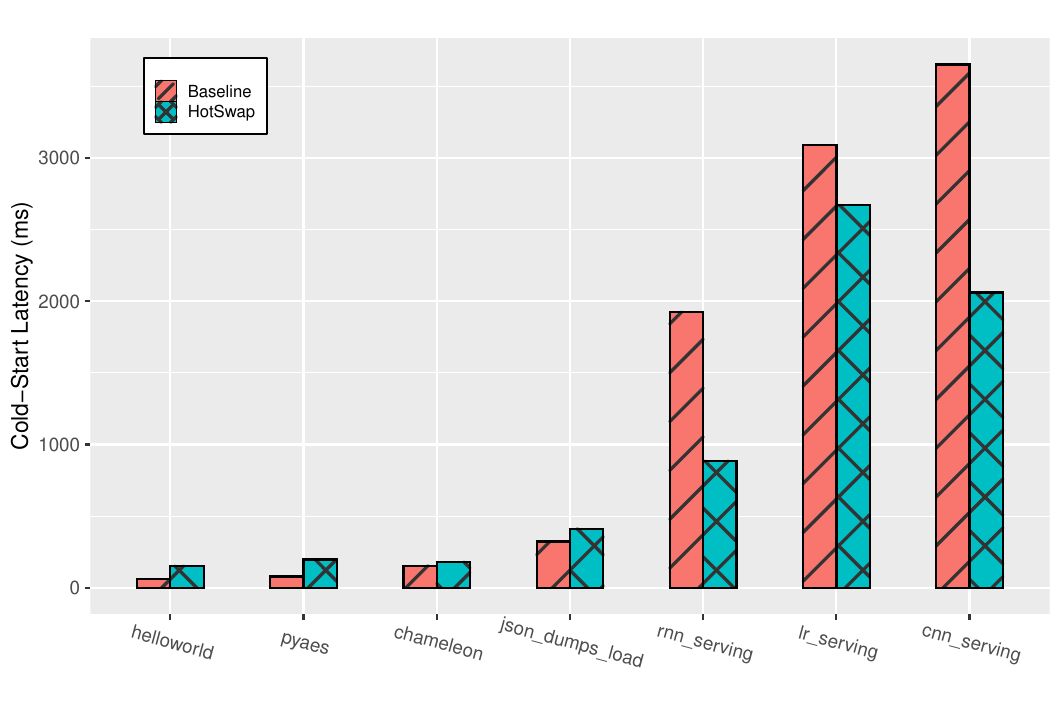}
  }
  \subfloat[Warm Start]
  {
    \label{fig:com_bas_warm}\includegraphics[width=0.45\textwidth]{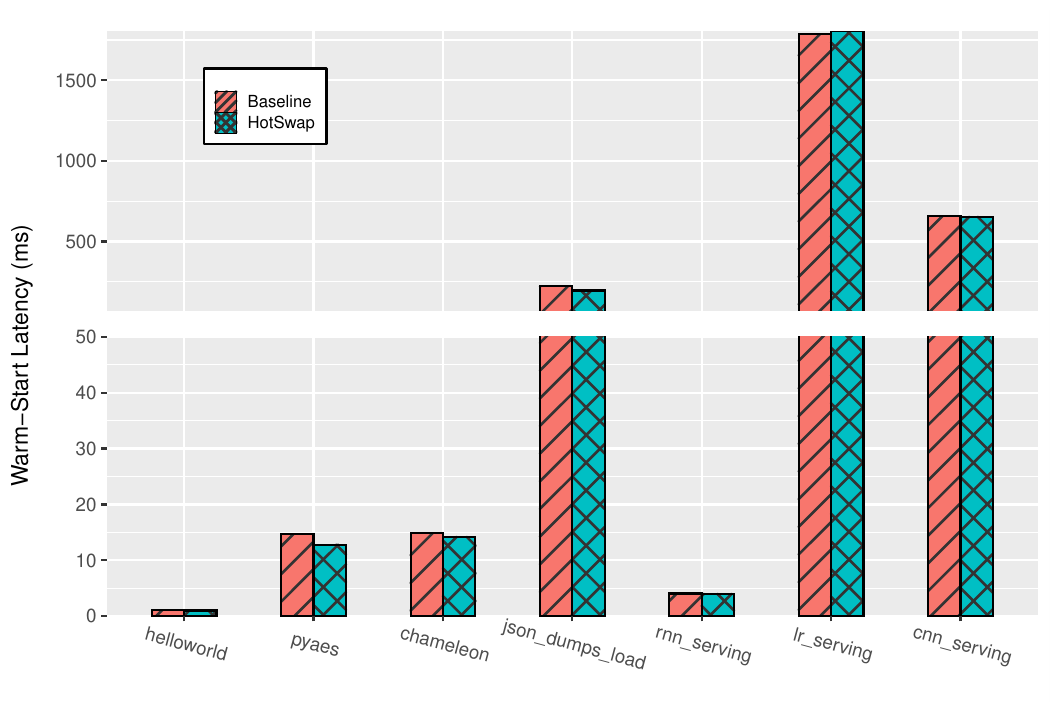} 
  }
  \caption{\textbf{Comparison of HotSwap versus Baseline.}  The software dependencies for each function are specified
  in Table~\ref{tab:AWS_performance} of Section~\ref{sec:AWS_performance_analysis}.  Note that \emph{lr\_serving}, \emph{cnn\_serving}, and \emph{rnn\_serving} all have much larger dependency images.  For this reason, HotSwap is shown to be much faster on cold starts than the Baseline in these cases, by a ratio of 1.2, 1.8, and 2.2.  Further, there is almost no degradation of performance during warm starts.}
  \label{fig:comprison_with_baseline}
\end{figure*}

\subsection{Implementation}
\label{sec:implementation}

We implement a prototype of HotSwap with AWS Lambda containers and CRIU (Checkpoint and Restore In Userspace)~\cite{criu}.
HotSwap uses AWS Lambda container templates (Python~3.9) to build serverless functions and collect performance results.
The dependency loader is embedded into the boot process of an AWS Lambda container.
When a serverless container receives an invocation, it first starts the dependency loader instead of the original ``bootstrap'' script.
Then the dependency loader loads the dependency image into the container and continually executes the AWS Lambda Runtime Interface~\cite{AWS_lambda_runtime} and the user code.

HotSwap is simulated on an AWS EC2 node instead of the online AWS Lambda environment.
This was necessary, since HotSwap is a provider-side implementation, and it is not possible to modify AWS Lambda.
Nevertheless, all experiments employ a standard AWS Lambda container and simply execute the container with the EC2 environment.
The EC2 environment was required because AWS Lambda does not allow the user to control the placement of serverless functions on nodes.  

The prototype modifies CRIU (Version~3.9) to implement live migration of dependency images. 
We add two new features supporting the design in Section~\ref{sec:dependencies_pool} and Section~\ref{sec:lazy_migration}.

First, we modified the ``\texttt{criu dump}'' command to support the dependency pool. 
This ensures that the ``\texttt{criu dump}'' command will always keep the live dependency image in memory (instead of being killed after migration) and managed by the dependency manager of HotSwap. 

Second, we modified the ``\texttt{criu pageserver}'' command to support the lazy restore policy described in Section~\ref{sec:lazy_migration}.
We extend the CRIU pageserver to accept multiple requests.
Note also that there is a ``\texttt{-{}-lazy page}'' parameter of the ``\texttt{criu pageserver}'' command.
However, this parameter migrates all pages upon the execution start.
We need to implement HotSwap's lazy restore policy: loading pages one by one.
Similarly, CRIU's \texttt{-{}-lazy page} parameter does not fully support HotSwap's bulk restore policy.
Some smaller modifications were also required for this policy.

We will release our configuration files and prototype code online as open source, to aid in reproducibility.

\section{Evaluation}
\label{sec:evaluation}

We first present the experimental setup in Section~\ref{sec:func_bench}, including the experimental platform and benchmarks for serverless functions.
The evaluation answers the following research questions:
\begin{itemize}
    \item \textbf{RQ1:} How well does HotSwap improve the cold start of different serverless functions? (Section~\ref{sec:basline_comp})
    \item \textbf{RQ2:} How does each HotSwap component affect cold starts across serverless functions? (Section~\ref{sec:HotSwap_breakdown})
    \item \textbf{RQ3:} How well does live migration in HotSwap improve the cold-start execution compared with standard migration using simple checkpoint images? (Section~\ref{sec:variants_comp})
    \item \textbf{RQ4:} How do different migration policies (lazy restore and bulk restore) work in HotSwap? (Section~\ref{sec:variants_comp})
    \item \textbf{RQ5:} How does HotSwap compare to PreBaking~\cite{silva2020prebaking, fireman2024prebaking} when sharing dependency images across multiple serverless functions in a real-world trace? (Section~\ref{sec:sharing_exp})
\end{itemize}

\subsection{Experimental Setup}
\label{sec:func_bench}
We use ``r5b.large'' AWS EC2 machines with two vCPUs, 16~GB memory, and AWS EBS 200~GB SSD, to conduct our experiments. 
The OS is Amazon Linux with Linux kernel 5.10.199, and the gcc version is 10.5.0.
As described in Section~\ref{sec:implementation}, an AWS EC2 node is required to simulate a provider-side HotSwap implementation, but standard AWS Lambda server containers are used in all experiments.
Moreover, all serverless containers are built based on the \linebreak[4]``public.ecr.aws/lambda/python:3.9'' template.
The serverless containers are started with ``\texttt{-{}-privileged}'' to allow the execution of CRIU.

All experiments use Python-based functions from the FunctionBench~\cite{kim2019functionbench,kim2019practical} serverless benchmark suite.
Recall that the middleware packages needed by these functions and other details are described in Table~\ref{tab:AWS_performance} in Section~\ref{sec:AWS_performance_analysis}.

Functions needing model parameters (namely, lr~\_serving, cnn~\_serving, rnn~\_serving) download their parameters from AWS S3 when executing cold starts. 
The dependency images created for each serverless function are based on Python and Python packages described in Table~\ref{tab:AWS_performance}.

The \emph{cold-start latency} discussed in this section refers to the times for: \emph{booting at runtime}; \emph{loading dependencies}; and \emph{executing the user functions}, unless otherwise specified.
To record the cold-start latency, we first start the serverless container in the worker node and wait for the invocations to eliminate the overhead of container creation.

\subsection{Comparing HotSwap and Baseline} 
\label{sec:basline_comp}
To answer \textbf{RQ1}, we evaluate the cold-start latency of HotSwap and the Baseline with all functions described in Section~\ref{sec:func_bench}. 
\emph{Baseline} refers to a traditional AWS Lambda container running offline.
The Baseline runs the same containers that were built for online AWS Lambda in Section~\ref{sec:AWS_performance_analysis} on an EC2 machine. 
In this experiment, HotSwap uses bulk restore (see Section~\ref{sec:lazy_migration}) because it has the best performance. These serverless containers directly boot the AWS Lambda Runtime Interface during cold starts and use a Python environment and packages pre-installed inside the serverless container.

Figure~\ref{fig:com_bas_cold} shows that HotSwap outperforms the Baseline on cold-start latency for serverless functions that require large dependencies.
HotSwap is slightly slower than Baseline for the lightweight serverless functions (namely, \emph{helloworld}, \emph{pyase}, \emph{chameleon}, and \emph{json\_dumps\_load}).
In production, HotSwap would be configured to avoid caching the trivial dependencies associated with these small serverless functions.
Nevertheless, HotSwap significantly outperforms Baseline for serverless functions that need larger packages (i.e., ``pandas'', ``NumPy'', ``scikit-learn'', ``torch'', and ``keras'').
\textbf{HotSwap is 1.2x, 1.8x, and 2.2x faster than Baseline for ``lr~\_serving'', ``cnn~\_serving'', and ``rnn~\_serving'', respectively.}

Figure~\ref{fig:com_bas_warm} shows the warm-start latency of HotSwap and Baseline.
For all six serverless functions, HotSwap has similar warm-start latencies with Baseline.

The cold start results show that HotSwap has a performance advantage only on certain serverless functions.
Specifically, HotSwap is designed to solve the large dependency-loading problem in the serverless cold start.
The lightweight functions only import standard and small Python packages.
So directly booting those functions is faster than using \linebreak[4]HotSwap.
The lightweight functions incur additional overhead because HotSwap needs to communicate with the dependency manager and migrate the dependency image into the containers, which increases some overhead.

The warm start results show that HotSwap's performance is almost the same as that of the Baseline containers. 
During warm starts, the user code is already executed, and necessary pages are already installed. 
HotSwap does not add overhead to the warm start of the same function. 
Furthermore, there is no difference between the execution times for regular software dependencies and for migrated dependencies.

\subsection{Components of Cold Start Execution Time}
\label{sec:HotSwap_breakdown}

To analyze HotSwap's cold-start performance and address \textbf{RQ2}, we decompose the cold-start time of HotSwap and Baseline into three parts.
Recall the architecture of \linebreak[4]HotSwap from Figure~\ref{fig:Metholodgy_system_overview}.
\emph{Communication} is the process by which the ``migration client'' communicates with the dependency manager and receives the process metadata.
The \emph{process metadata} is a special data structure associated with checkpoints that support CRIU's lazy migration protocol (see Section~\ref{sec:dependencies_pool}).
\emph{Migration} is the process in which the dependency loader executes until the start of the user-specific handler function.
The definitions of \emph{Execution}, \emph{Boot}, and \emph{Dependency Initialization} are the same as those in Section~\ref{sec:AWS_performance_analysis}.

\begin{figure}[ht]
    \centering
    \includegraphics[width=0.48\textwidth]{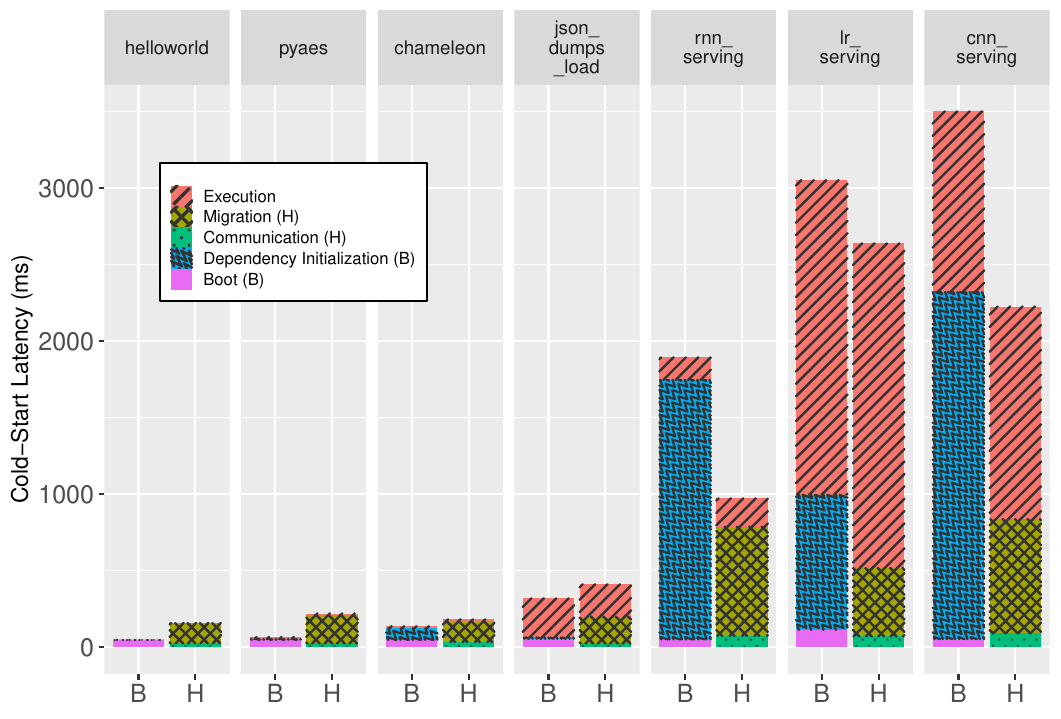}
    \caption{\textbf{Cold-start latency breakdown for HotSwap (H) and Baseline (B).} The performance gain of HotSwap arises because the migration is faster than the software dependency initialization from scratch.} 
    \label{fig:CS_breakdown}
\end{figure}

Figure~\ref{fig:CS_breakdown} shows the breakdown of cold-start latency for HotSwap.
The communication time is almost constant for all tested serverless functions, ranging from 25~ms to 94~ms according to the sizes of process metadata. 
Table~\ref{tab:file_size} shows the sizes of process metadata for dependency images for each function.
The size of process metadata is much smaller than the checkpoint image of the entire dependency.

\begin{table}[htbp]
\centering
\caption{The size of data that needs to be transferred from the worker node to the function container of HotSwap and HotSwap without Page Server.}
\label{tab:file_size}
\begin{tabular}{lll}
\hline
\multicolumn{1}{c}{\multirow{2}{*}{\begin{tabular}[c]{@{}c@{}} Functions\end{tabular}}} & \multicolumn{1}{c}{\multirow{2}{*}{\begin{tabular}[c]{@{}c@{}}Metadata (MB)\end{tabular}}} & \multicolumn{1}{c}{\multirow{2}{*}{\begin{tabular}[c]{@{}c@{}}Image (MB)\end{tabular}}} \\
\multicolumn{1}{c}{}                                                                                 & \multicolumn{1}{c}{}                                                                                       & \multicolumn{1}{c}{}                                                                                      \\ \hline
helloworld                                                                                           & 0.88                                                                                                       & 8.1                                                                                                       \\
json\_dumps\_load                                                                                    & 1.2                                                                                                        & 16                                                                                                        \\
pyase                                                                                                & 0.88                                                                                                       & 8.3                                                                                                       \\
chameleon                                                                                            & 0.88                                                                                                       & 8.9                                                                                                       \\
lr\_serving                                                                                          & 5.6                                                                                                        & 79                                                                                                        \\
cnn\_serving                                                                                         & 15                                                                                                         & 190                                                                                                       \\
rnn\_serving                                                                                         & 12                                                                                                         & 200                                                                                                       \\ \hline
\end{tabular}
\end{table}

Compared with the dependency-loading time of Baseline, HotSwap has a faster migration time in the case of functions that have larger dependencies.
\textbf{HotSwap is 2.2x, 3.2x, and 2.5x faster than Baseline for booting dependencies for ``lr~\_serving'', ``cnn~\_serving'' and ``rnn~\_serving'', respectively.}
That explains how HotSwap helps with the cold-start time.
Since all memory pages are pre-initialized in the dependency image, user functions directly use them after migration.
Within HotSwap, the migration time dominates the cold start because migration is the key phase that transfers the dependency image that is needed by the user function.

\subsection{Ablation Study} 
\label{sec:variants_comp}

This section evaluates the contribution of the different optimization techniques used by HotSwap in cold starts: live migration (\textbf{RQ3}); and different migration policies (\textbf{RQ4}). Since HotSwap shows performance improvements only for functions with larger dependencies, we focus on the results for ``lr\_serving,'' ``cnn\_serving,'' and ``rnn\_serving''.

\begin{table*}[htbp]
\centering
\caption{Comparison of cold-start and warm-start results with: HotSwap bulk restore, HotSwap Lazy Restore; HotSwap without Page Server; and HotSwap without Lazy Migration.}
\label{tab:Prototypes_Comparison}
\begin{tabular}{lrrr|rrr}
\hline
\multirow{2}{*}{Prototypes} & \multicolumn{3}{c|}{Cold Start (s)}                                                & \multicolumn{3}{c}{Warm Start (s)}                                                \\ \cline{2-7} 
                            & lr\_serving & cnn\_serving &rnn\_serving & lr\_serving & cnn\_serving &rnn\_serving \\ \hline
HotSwap Bulk Restore                & \textbf{2.67}                 & \textbf{2.06}                  & 0.89                   & 1.80                 & 0.65                   & 0.004                     \\
HotSwap Lazy Restore               & 2.93                 & 3.29                  & \textbf{0.76}                   & 2.39                 & 2.71                   & 0.004                     \\
w/o Page Server             & 3.00                 & 2.14                  & 0.91                   & 1.82                  & 0.68                    & 0.004                     \\
w/o Lazy Migration               & 3.09                 & 2.13                  & 0.89                   & 1.81                  & 0.65                        & 0.004                        \\ \hline
\end{tabular}
\end{table*}

Two migration policies of HotSwap from Section~\ref{sec:lazy_migration} are evaluated:
\begin{description}
    \item[\emph{HotSwap Bulk Restore}:] The page server copies all memory pages in parallel after the first ``page fault''.
    \item[\emph{HotSwap Lazy Restore}:] The page server copies only the accessed pages that trigger a ``page fault''.
\end{description}

Two variants of HotSwap are created as follows:
\begin{description} 
    \item[\emph{HotSwap without Page Server}:] Prepares a regular checkpoint image for the dependency, copies it to the user container during cold starts, and restores it.
    \item[\emph{HotSwap without Lazy Migration}:] Copies all memory pages \linebreak[4]through the page server inside the user container before the execution of user code.
\end{description}
Table~\ref{tab:Prototypes_Comparison} shows the cold-start and warm-start latencies for HotSwap and its variants. 
HotSwap with bulk restore has the best latency for ``lr\_serving'' and ``cnn\_serving'', while HotSwap with lazy restore outperforms the others for \linebreak[4]``rnn\_serving''. 
The cold-start performance gap between bulk and lazy restore depends on page faults, as each fault in lazy restore pauses execution to wait for the page server.

For HotSwap without a Page Server, we copy a relatively large image to the user container and restore all memory pages before execution.  
The overhead of copying is related to the checkpoint sizes listed in Table~\ref{tab:file_size}.

For HotSwap without Lazy Migration, the cold-start time is slower than HotSwap because it needs to transfer all the memory pages through the network before the start of function execution.
Note that for ``lr~\_serving'', HotSwap without Lazy Migration is even slower than HotSwap without page server. 
This is because ``lr~\_serving'' has a smaller checkpoint image size than the other functions.

All three prototypes (excluding Lazy Restore) have similar warm-start times, demonstrating that these migration methods have negligible impact on warm-start latency.
For Lazy Restore, the performance of warm starts decreases because ``lr\_serving'' and ``cnn\_serving'' trigger extra page faults during the first warm start. 

Based on these experiments, we believe bulk restore is more suitable in practice since the provider usually does not know how many page faults the target program will trigger. 

\subsection{Case Study: Dependency Image Sharing Across Serverless Functions} 
\label{sec:sharing_exp} 

In this section, we conduct simulation experiments to demonstrate the benefits of HotSwap in sharing dependency images across different serverless functions, addressing \textbf{RQ5}. 
We use real-world serverless function traces from Azure~\cite{shahrad2020serverless}, previously discussed in Section~\ref{sec:invocation_pattern_vs_cold_start} of this paper, as the basis for our tests. 
We modified the ``rnn\_serving'' function by adding new code (utilizing NumPy and Torch) to create 10 different serverless functions. 
For each function, we assigned random traces and measured latency over a two-week period. 
The keep-alive time was set to $15$ minutes. 
Since HotSwap is designed to optimize performance for less frequently used functions, we tested it under various invocation frequencies.

This experiment compares HotSwap with a checkpoint image-based method, Prebaking~\cite{silva2020prebaking, fireman2024prebaking}.
Prebaking is chosen as a well-known system that employs CRIU-based checkpoint/restore to accelerate serverless cold starts.
We implemented Prebaking as described in their paper, by checkpointing the entire \emph{previously warm serverless container} using CRIU~\cite{criu} to create a function image, which is then restored during a cold start. 
To enhance fairness, we store all prebaking images in memory.
Note that the cold start latency measured in this experiment \textbf{includes the container creation overhead for both methods.}

\begin{figure}[ht]
    \centering
    \includegraphics[width=0.48\textwidth]{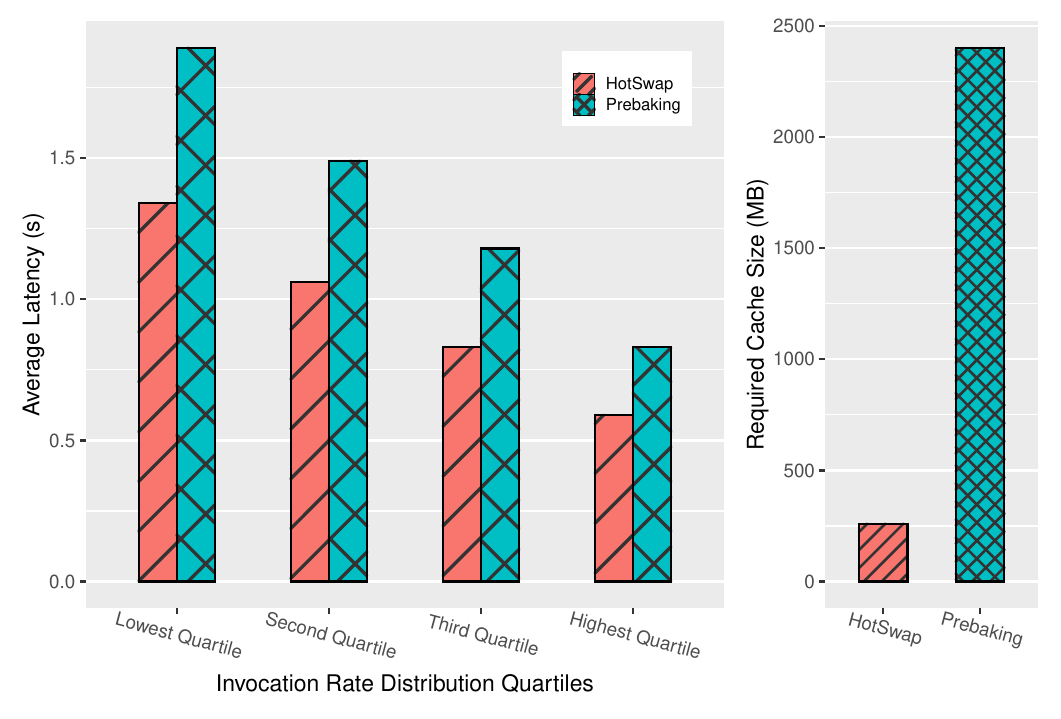} 
    \caption{\textbf{Comparison of average latency and required cache/memory between HotSwap and Prebaking\cite{silva2020prebaking, fireman2024prebaking} under Azure Function traces \cite{shahrad2020serverless}.} The experiment involves 10 different serverless functions, all using the same dependency image ( NumPy + Torch). Each function is assigned a real-world invocation trace from Azure. The left figure shows latency across different invocation rate groups, ranging from the lowest to the highest quartile.}
    \label{fig:Optimization_cost_comparison}
\end{figure}

Figure~\ref{fig:Optimization_cost_comparison} shows the average latency and required cache for HotSwap and Prebaking.
The average latency is calculated from all invocations of the ten different functions.
To evaluate different invocation rate scenarios, we categorized the Azure traces into quartiles based on invocation rate. 
The ``Lowest Quartile'' represents experiments on function invocation traces with rates below the 25th percentile, while the other three groups correspond to ``25-50\%'', ``50-75\%'', and above the 75th percentile, respectively.

The latency results indicate that HotSwap outperforms Prebaking by 1.4x across all trace groups. 
This improvement is due to HotSwap’s superior cold-start performance. 
The average latency decreases as the invocation rate increases, resulting in more warm starts. 
Notably, the 75th percentile of invocation rates from Azure traces~\cite{shahrad2020serverless} is 0.04 calls per minute.
This demonstrates that the majority of serverless functions are, in fact, used infrequently.

Furthermore, the provider requires only 260 MB of memory to warm up the 10 different functions when using HotSwap. 
In contrast, Prebaking requires the provider to maintain 10 checkpoint images (approximately 2.4 GB) in the cache for the same functions. 
Therefore, warming up the less frequently used functions with HotSwap is more cost-effective than Prebaking. 
This benefit can increase further if there are more functions that share the same dependency image.

Although function-specific tuning methods, like vHive~\cite{ustiugov2021benchmarking} and Catalyzer~\cite{du2020catalyzer}, achieve faster container restore, the provider must create and keep all function images in cache to optimize the less frequently used serverless functions.
Thus, HotSwap provides a practical method for providers to optimize cold starts for skewed workloads at a reasonable cost.

\section{Related Work}
\label{sec:related_work}
\noindent \textbf{Pre-Warm.} One straightforward way to optimize cold starts is to prevent them from happening. 
Prior research has explored various approaches such as adjusting keep-alive policies, leveraging application-level composition~\cite{bermbach2020using}, mining invocation patterns~\cite{shen2021defuse}, and applying machine learning models~\cite{lu2024smiless, kumari2022mitigating, agarwal2024demand},
in order to predict function usage and proactively pre-warm containers~\cite{roy2022icebreaker, fuerst2021faascache, shahrad2020serverless}.
However, these strategies often lead to resource inefficiency due to inaccurate predictions, especially in dynamic production environments.
HotSwap takes a different approach by directly reducing cold-start overhead instead of relying on speculative pre-warming.

\noindent \textbf{Function-Specific Optimization.} A common method for reducing cold-start latency is leveraging Checkpoint/Restore (C/R) techniques such as CRIU~\cite{criu}. 
Prebaking~\cite{silva2020prebaking, fireman2024prebaking} applies CRIU to checkpoint entire serverless functions, significantly speeding up start-up across runtimes.
Catalyzer~\cite{du2020catalyzer} improves sandbox restoration by efficiently resuming serverless functions across sandboxes. 
MITOSIS~\cite{wei2023no} extends this approach by introducing remote sandbox forking using RDMA. 

Further extending this idea, Ustiugov et al.~\cite{ustiugov2021benchmarking} propose a ``hot page record-and-prefetch'' technique to accelerate sandbox restoration and function execution. 
SEUSS~\cite{cadden2020seuss} takes a unikernel-based approach, restoring function snapshots while avoiding resource-heavy initialization. 
Fasslight~\cite{liu2023faaslight} optimizes code loading latency by loading only essential code. 
Although effective for high-frequency serverless functions, these methods may not suit skewed workloads in production, where checkpointing and analysis overhead can outweigh the benefits.  
Furthermore, a recent user-perspective study~\cite{dos2023alternative} suggests that Node.js performs better than Java for infrequently invoked functions.  
However, providers have no control over which language users choose to implement their functions.

\noindent \textbf{Cross-Function Optimization and Dependency Loading.} 
Cross-function optimizations improve performance by sharing resources across functions.  
Pagurus~\cite{li2021pagurus} reduces cold-start overhead via container reuse, while HotC~\cite{suo2021tackling} maintains a pool of preloaded runtimes for reuse based on function inputs.

Software dependency loading is a key factor in cold-start latency.
SOCK~\cite{oakes2018sock} uses pre-initialized Zygote containers and multi-tier caching to speed up Python library loading.
(In fact, SOCK uses a generalization of the original Android-based Zygotes.)
FlashCube~\cite{lin2021flashcube} builds containers from prebuilt components to reduce initialization time.
Photons~\cite{dukic2020photons} co-locates function instances in a shared runtime to enable dependency reuse.
TrEnv~\cite{huang2024trenv} leverages remote memory pools via CXL/RDMA to share execution environments.
SLIMSTART~\cite{ebrahimi2024cold} removes infrequently used libraries through runtime profiling and automated transformation.
HotSwap aims to introduce a more lightweight and easily implementable solution to reduce dependency-related overhead.

\noindent \textbf{Infrastructure and Storage Systems.} 
Infrastructure optimization can further reduce cold-start latency by improving storage and caching mechanisms. 
Firecracker~\cite{agache2020firecracker}, a micro-VM developed by AWS Lambda, leverages the Linux Kernel’s KVM virtualization to provide lightweight virtual machines, significantly reducing boot times and improving context-switching performance. 
AWS also optimizes serverless container sharing by deduplicating data chunks across functions, revealing that 80\% of uploaded containers contain redundant data~\cite{brooker2023demand, meyer2012study, xia2016comprehensive}. 
These infrastructure optimizations are complementary to HotSwap, as they focus on different aspects of serverless performance improvements.

\section{Limitations}
\label{sec:limit}
\noindent 
\textbf{HotSwap is specifically designed for to be active during the dependency loading phase.}  
HotSwap targets the overhead associated with dependency loading and is most effective when this overhead is substantial.  
For functions with only trivial dependencies, using HotSwap to accelerate dependency loading may not be necessary.  
From the provider's perspective, this also means that it is not required to cache all dependencies, only those that are popular and large.

\noindent 
\textbf{Evaluation on Other Runtimes and Languages.}  
In this paper, we evaluate HotSwap only on the Python runtime.  
We chose Python because it is one of the most popular languages for serverless functions, particularly in data processing, automation, and machine learning tasks.  
Moreover, many widely used Python libraries are large and contribute significantly to cold start overhead.  
However, since HotSwap relies on process-level migration, it could, in theory, be applied to other runtimes and languages as well—especially in scenarios where dependency loading overhead is a major concern.

\noindent 
\textbf{Centralized Dependency Management.}  
The Dependency Manager introduced in Section~\ref{sec:design_overview} is centralized in our prototype design and may become a bottleneck when handling numerous migration requests.  
In practice, cloud providers can implement HotSwap as a distributed caching system with multiple Dependency Managers operating within a cluster.

\section{Conclusion}
\label{sec:conclusion}
In conclusion, this paper presents HotSwap, a cross-function optimization designed to accelerate dependency loading in serverless cold starts.
HotSwap provides a strong benefit in cases of skewed workloads, with many infrequently used serverless functions (the problem of a ``long tail'' in the distribution).
It primarily optimizes the dependency loading phase, a major bottleneck in current production clouds.
By enabling a shared pool of dependency images across multiple compute hosts, HotSwap allows providers to efficiently manage commonly used middleware.
We implemented a HotSwap prototype based on offline AWS Lambda containers, and our evaluation demonstrates that it significantly reduces dependency loading overhead and warm-up costs across various serverless functions.

\medskip
\noindent 
\textbf{Acknowledgment.}  
We thank our shepherd, Yehia Elkhatib, and the anonymous reviewers for their constructive feedback. This work is supported in part by NSF Awards 1910601 and 2124897.

\bibliographystyle{ieeetr}
\bibliography{references}

\begin{thebibliography}{10}

\bibitem{AWSLambda}
``{AWS} {L}ambda.'' \url{https://aws.amazon.com/lambda/}.
\newblock Accessed: 2024-07-15.

\bibitem{AzureFunction}
``Microsoft {A}zure {F}unction.'' \url{https://azure.microsoft.com/en-us/products/functions/}.
\newblock Accessed: 2024-07-15.

\bibitem{googlecloud}
``Google cloud serverless.'' \url{https://cloud.google.com/serverless}.
\newblock Accessed: 2024-07-15.

\bibitem{AlibabaFunction}
``Alibaba {F}unction {C}omputing.'' \url{https://www.alibabacloud.com/product/function-compute}.
\newblock Accessed: 2024-07-15.

\bibitem{ristov2022colder}
S.~Ristov, C.~Hollaus, and M.~Hautz, ``Colder than the warm start and warmer than the cold start! experience the spawn start in faas providers,'' in {\em Proceedings of the 2022 Workshop on Advanced tools, programming languages, and PLatforms for Implementing and Evaluating algorithms for Distributed systems}, pp.~35--39, 2022.

\bibitem{manner2018cold}
J.~Manner, M.~Endre{\ss}, T.~Heckel, and G.~Wirtz, ``Cold start influencing factors in function as a service,'' in {\em 2018 IEEE/ACM International Conference on Utility and Cloud Computing Companion (UCC Companion)}, pp.~181--188, IEEE, 2018.

\bibitem{ebrahimi2024cold}
A.~Ebrahimi, M.~Ghobaei-Arani, and H.~Saboohi, ``Cold start latency mitigation mechanisms in serverless computing: taxonomy, review, and future directions,'' {\em Journal of systems architecture}, p.~103115, 2024.

\bibitem{fuerst2021faascache}
A.~Fuerst and P.~Sharma, ``Faascache: keeping serverless computing alive with greedy-dual caching,'' in {\em Proceedings of the 26th ACM International Conference on Architectural Support for Programming Languages and Operating Systems}, pp.~386--400, 2021.

\bibitem{shen2021defuse}
J.~Shen, T.~Yang, Y.~Su, Y.~Zhou, and M.~R. Lyu, ``Defuse: A dependency-guided function scheduler to mitigate cold starts on faas platforms,'' in {\em 2021 IEEE 41st International Conference on Distributed Computing Systems (ICDCS)}, pp.~194--204, IEEE, 2021.

\bibitem{roy2022icebreaker}
R.~B. Roy, T.~Patel, and D.~Tiwari, ``Icebreaker: Warming serverless functions better with heterogeneity,'' in {\em Proceedings of the 27th ACM International Conference on Architectural Support for Programming Languages and Operating Systems}, pp.~753--767, 2022.

\bibitem{shahrad2020serverless}
M.~Shahrad, R.~Fonseca, I.~Goiri, G.~Chaudhry, P.~Batum, J.~Cooke, E.~Laureano, C.~Tresness, M.~Russinovich, and R.~Bianchini, ``Serverless in the wild: Characterizing and optimizing the serverless workload at a large cloud provider,'' in {\em 2020 USENIX annual technical conference (USENIX ATC 20)}, pp.~205--218, 2020.

\bibitem{wang2018peeking}
L.~Wang, M.~Li, Y.~Zhang, T.~Ristenpart, and M.~Swift, ``Peeking behind the curtains of serverless platforms,'' in {\em 2018 USENIX annual technical conference (USENIX ATC 18)}, pp.~133--146, 2018.

\bibitem{silva2020prebaking}
P.~Silva, D.~Fireman, and T.~E. Pereira, ``Prebaking functions to warm the serverless cold start,'' in {\em Proceedings of the 21st International Middleware Conference}, pp.~1--13, 2020.

\bibitem{fireman2024prebaking}
D.~Fireman, P.~Silva, T.~E. Pereira, L.~Mafra, and D.~Valadares, ``Prebaking runtime environments to improve the faas cold start latency,'' {\em Future Generation Computer Systems}, vol.~155, pp.~287--299, 2024.

\bibitem{du2020catalyzer}
D.~Du, T.~Yu, Y.~Xia, B.~Zang, G.~Yan, C.~Qin, Q.~Wu, and H.~Chen, ``Catalyzer: Sub-millisecond startup for serverless computing with initialization-less booting,'' in {\em Proceedings of the Twenty-Fifth International Conference on Architectural Support for Programming Languages and Operating Systems}, pp.~467--481, 2020.

\bibitem{cadden2020seuss}
J.~Cadden, T.~Unger, Y.~Awad, H.~Dong, O.~Krieger, and J.~Appavoo, ``{SEUSS}: Skip redundant paths to make serverless fast,'' in {\em Proceedings of the Fifteenth European Conference on Computer Systems}, pp.~1--15, 2020.

\bibitem{ustiugov2021benchmarking}
D.~Ustiugov, P.~Petrov, M.~Kogias, E.~Bugnion, and B.~Grot, ``Benchmarking, analysis, and optimization of serverless function snapshots,'' in {\em Proceedings of the 26th ACM International Conference on Architectural Support for Programming Languages and Operating Systems}, pp.~559--572, 2021.

\bibitem{wei2023no}
X.~Wei, F.~Lu, T.~Wang, J.~Gu, Y.~Yang, R.~Chen, and H.~Chen, ``No provisioned concurrency: Fast {RDMA}-codesigned remote fork for serverless computing,'' in {\em 17th USENIX Symposium on Operating Systems Design and Implementation (OSDI 23)}, pp.~497--517, 2023.

\bibitem{oakes2018sock}
E.~Oakes, L.~Yang, D.~Zhou, K.~Houck, T.~Harter, A.~Arpaci-Dusseau, and R.~Arpaci-Dusseau, ``{SOCK}: {R}apid task provisioning with serverless-optimized containers,'' in {\em 2018 USENIX Annual Technical Conference USENIX (ATC'18))}, pp.~57--70, 2018.

\bibitem{huang2024trenv}
J.~Huang, M.~Zhang, T.~Ma, Z.~Liu, S.~Lin, K.~Chen, J.~Jiang, X.~Liao, Y.~Shan, N.~Zhang, {\em et~al.}, ``Trenv: Transparently share serverless execution environments across different functions and nodes,'' in {\em Proceedings of the ACM SIGOPS 30th Symposium on Operating Systems Principles}, pp.~421--437, 2024.

\bibitem{lin2021flashcube}
Z.~Lin, K.-F. Hsieh, Y.~Sun, S.~Shin, and H.~Lu, ``Flashcube: Fast provisioning of serverless functions with streamlined container runtimes,'' in {\em Proceedings of the 11th Workshop on Programming Languages and Operating Systems}, pp.~38--45, 2021.

\bibitem{brooker2023demand}
M.~Brooker, M.~Danilov, C.~Greenwood, and P.~Piwonka, ``On-demand container loading in {AWS} {L}ambda,'' in {\em 2023 USENIX Annual Technical Conference (USENIX ATC 23)}, pp.~315--328, 2023.

\bibitem{agache2020firecracker}
A.~Agache, M.~Brooker, A.~Iordache, A.~Liguori, R.~Neugebauer, P.~Piwonka, and D.-M. Popa, ``Firecracker: {L}ightweight virtualization for serverless applications,'' in {\em 17th USENIX symposium on networked systems design and implementation (NSDI 20)}, pp.~419--434, 2020.

\bibitem{kim2019functionbench}
J.~Kim and K.~Lee, ``Function{B}ench: A suite of workloads for serverless cloud function service,'' in {\em 2019 IEEE 12th International Conference on Cloud Computing (CLOUD)}, pp.~502--504, IEEE, 2019.

\bibitem{kim2019practical}
J.~Kim and K.~Lee, ``Practical cloud workloads for serverless {FaaS},'' in {\em Proceedings of the ACM Symposium on Cloud Computing}, pp.~477--477, 2019.

\bibitem{AWS_lambda_runtime}
``{AWS} {L}ambda runtime interface client for {P}ython.'' \url{https://github.com/aws/aws-lambda-python-runtime-interface-client/releases/tag/2.0.8}.
\newblock Accessed: 2024-01-13.

\bibitem{criu}
``{CRIU}: checkpoint and restore in userspace..'' \url{https://criu.org/}.
\newblock Accessed: 2024-07-15.

\bibitem{bermbach2020using}
D.~Bermbach, A.-S. Karakaya, and S.~Buchholz, ``Using application knowledge to reduce cold starts in faas services,'' in {\em Proceedings of the 35th annual ACM symposium on applied computing}, pp.~134--143, 2020.

\bibitem{lu2024smiless}
C.~Lu, H.~Xu, Y.~Li, W.~Chen, K.~Ye, and C.~Xu, ``Smiless: Serving dag-based inference with dynamic invocations under serverless computing,'' in {\em SC24: International Conference for High Performance Computing, Networking, Storage and Analysis}, pp.~1--17, IEEE, 2024.

\bibitem{kumari2022mitigating}
A.~Kumari, B.~Sahoo, and R.~K. Behera, ``Mitigating cold-start delay using warm-start containers in serverless platform,'' in {\em 2022 IEEE 19th India Council International Conference (INDICON)}, pp.~1--6, IEEE, 2022.

\bibitem{agarwal2024demand}
S.~Agarwal, M.~A. Rodriguez, and R.~Buyya, ``On-demand cold start frequency reduction with off-policy reinforcement learning in serverless computing,'' in {\em International Conference on Computing, Intelligence and Data Analytics}, pp.~1--24, Springer, 2024.

\bibitem{liu2023faaslight}
X.~Liu, J.~Wen, Z.~Chen, D.~Li, J.~Chen, Y.~Liu, H.~Wang, and X.~Jin, ``Faaslight: General application-level cold-start latency optimization for function-as-a-service in serverless computing,'' {\em ACM Transactions on Software Engineering and Methodology}, vol.~32, no.~5, pp.~1--29, 2023.

\bibitem{dos2023alternative}
P.~O.~F. Dos~Santos, H.~J. de~Moura~Costa, V.~R. Leithardt, and P.~J.~S. Ferreira, ``An alternative to faas cold start latency of low request frequency applications,'' in {\em 2023 3rd International Conference on Electrical, Computer, Communications and Mechatronics Engineering (ICECCME)}, pp.~1--6, IEEE, 2023.

\bibitem{li2021pagurus}
Z.~Li, Q.~Chen, and M.~Guo, ``Pagurus: Eliminating cold startup in serverless computing with inter-action container sharing,'' {\em arXiv preprint arXiv:2108.11240}, 2021.

\bibitem{suo2021tackling}
K.~Suo, J.~Son, D.~Cheng, W.~Chen, and S.~Baidya, ``Tackling cold start of serverless applications by efficient and adaptive container runtime reusing,'' in {\em 2021 IEEE International Conference on Cluster Computing (CLUSTER)}, pp.~433--443, IEEE, 2021.

\bibitem{dukic2020photons}
V.~Dukic, R.~Bruno, A.~Singla, and G.~Alonso, ``Photons: Lambdas on a diet,'' in {\em Proceedings of the 11th ACM Symposium on Cloud Computing}, pp.~45--59, 2020.

\bibitem{meyer2012study}
D.~T. Meyer and W.~J. Bolosky, ``A study of practical deduplication,'' {\em ACM Transactions on Storage (ToS)}, vol.~7, no.~4, pp.~1--20, 2012.

\bibitem{xia2016comprehensive}
W.~Xia, H.~Jiang, D.~Feng, F.~Douglis, P.~Shilane, Y.~Hua, M.~Fu, Y.~Zhang, and Y.~Zhou, ``A comprehensive study of the past, present, and future of data deduplication,'' {\em Proceedings of the IEEE}, vol.~104, no.~9, pp.~1681--1710, 2016.

\end{thebibliography}
\end{document}